\documentclass[showkeys,preprint]{revtex4}
\usepackage{graphicx,booktabs,array}
\usepackage{amsmath}
\usepackage{amssymb}

\usepackage{graphicx,mathdots}
\usepackage{color}
\usepackage{booktabs}
\usepackage{mathtools,tabularx}

\usepackage{makecell}
\setcellgapes{3pt}

\begin{document}

\setcounter{page}{1}


\title{Dyonic rotating cosmological black hole surrounded by quintessence}

\author{M. D. de Oliveira\footnote{Corresponding author. Email: dalpra.matheus@gmail.com} and Alexandre G. M. Schmidt}
\email{Telephone: (+55) 24 3076 8959, FAX: (+55) 24 3076 8870}
\affiliation{Instituto de Ci\^encias Exatas, Universidade Federal Fluminense,\\ 
27213-145 Volta Redonda - RJ, Brazil}


\begin{abstract}
In this work, we constructed a new metric describing a rotating cosmological black hole with electric and magnetic charges in the presence of quintessence dark energy, where the cosmological constant and quintessence are associated with external matter. Starting from a Schwarzschild-type metric and considering all energy contributions in the function $f(r)$, we introduced rotation through the Newman--Janis algorithm using a unique and general complexification rule, avoiding possible ambiguities. Assuming that the metric satisfies the Einstein field equations with matter, we calculate the event horizon condition, the total stress--energy tensor $T_{\mu\nu}$, and the Kretschmann and Ricci scalars. We also analyze the influence of the cosmological constant $\Lambda$ and quintessence parameter $\alpha$ on the event horizon and ergosphere. In the limit $\alpha \rightarrow 0$, the scalar curvature differs from the vacuum result, $R \neq -4\Lambda$, while the singularity region remains unchanged. Finally, we investigate the angular and rotational velocities of a test particle, as well as the energy conditions, through the energy density and pressures required for the existence of this black hole solution.
\end{abstract}


\maketitle

\section{Introduction}

The study of black holes remains one of the central topics in General Relativity and modern gravitational physics. Since the Schwarzschild solution for a static and spherically symmetric spacetime \cite{gravitation}, several extensions have been proposed in order to describe more realistic astrophysical systems, including rotation, electric charge, magnetic charge, cosmological constant, and different matter sources. The Kerr solution generalized the Schwarzschild geometry to rotating black holes \cite{kerr}, while the Kerr--Newman metric included the presence of electric charge \cite{newman}. Subsequently, the introduction of the cosmological constant led to Kerr--Newman--AdS-type solutions, which play an important role in black hole thermodynamics, quantum gravity, and the AdS/CFT correspondence \cite{carter0,gibbons}.

In recent decades, the discovery of the accelerated expansion of the Universe through supernova observations \cite{riess,perlmutter} motivated the study of dark energy models. Among the most important candidates are the cosmological constant and quintessence fields, characterized by the energy term $\alpha r^{-1-3\omega}$, where $\alpha$ and $-1<\omega<-1/3$ are real parameters \cite{ratra,copeland}. In this context, Kiselev obtained an exact solution of the Einstein equations describing a black hole surrounded by quintessence \cite{kiselev}. This solution opened the possibility of investigating the influence of dark energy on black hole geometries and their physical properties. Subsequently, several works studied black holes surrounded by quintessence, including analyses of horizon structure, thermodynamics, geodesics, Hawking radiation, quasinormal modes, and gravitational lensing \cite{chen,toshmatov1,fernando,ghosh}. Since realistic astrophysical black holes are expected to possess angular momentum, rotating solutions became particularly relevant. In this scenario, the Newman--Janis algorithm (NJA) \cite{newmanjannis1} became one of the most widely used methods for generating rotating metrics from static and spherically symmetric spacetimes. Using this method, Ghosh constructed a rotating black hole surrounded by quintessence \cite{ghosh}, while Xu and Wang later generalized the solution to the Kerr--Newman--AdS black hole surrounded by quintessence \cite{xu}. More recently, different authors have investigated rotating black holes in the presence of quintessence and cosmological fluids in contexts involving black hole shadows, accretion disks, particle dynamics, and observational signatures \cite{alam,kumar,atika}.

Furthermore, the search for new black hole solutions in different gravitational scenarios has intensified in recent years, motivated by studies involving regular black holes, nonlinear electrodynamics, modified gravity theories, magnetic charges, anisotropic fluids, and dark energy effects \cite{bardeen1,ayonbeato,nojiri,toshmatov2,kasuya}. In particular, dyonic black holes, which simultaneously possess electric and magnetic charges, have attracted considerable attention due to their rich geometrical and thermodynamical structure \cite{dadhich,pradhan}.

Despite these advances, most known Kerr--Newman--AdS-type solutions treat the cosmological constant only as a vacuum geometric contribution in the Einstein field equations. However, from a physical perspective, both the cosmological constant and quintessence may be interpreted as effective matter sources distributed throughout spacetime. This interpretation directly modifies the stress--energy tensor and, consequently, changes the resulting geometry. Therefore, the construction of new black hole solutions that explicitly include external matter associated with both the cosmological constant and quintessence becomes relevant in relativistic cosmology and gravitational physics.

In this work, we construct a new solution describing a rotating cosmological black hole carrying both electric and magnetic charges and surrounded by quintessence, explicitly considering external matter associated with the cosmological constant and quintessential dark energy. We investigate the geometrical properties of this new metric through the horizon structure, ergosphere, scalar curvature, as well as the Einstein and stress--energy tensors. In addition, the angular and rotational velocities obtained from test particle dynamics and the energy conditions in the locally non-rotating frame (LNRF) are also analyzed.

The outline of this paper is as follows: In Section II, we apply the Newman--Janis algorithm to introduce rotation into a charged cosmological black hole surrounded by quintessence. After obtaining the corresponding line element, we also investigate the possible event horizons and the ergosphere structure. In Section III, we calculate the electromagnetic tensor and the total stress--energy tensor satisfying the Einstein field equations in the presence of matter. We then examine the existence of singularities, as well as the angular and rotational velocities, angular momentum, and mechanical energy of a test particle. Finally, we analyze the energy conditions required for the existence of this black hole solution. In Section IV, we present our conclusions.

\section{Newman-Janis algorithm applied to the Schwarzschild-type metric}

In this section, we will calculate the rotating charged cosmological black hole surrounded by quintessence through the Newman–Janis algorithm (NJA). Thus, we begin by writing the general Schwarzschild-type metric, considered as a static spherically symmetric seed metric, given by
\begin{equation}\label{metricageralf}
	ds^2 = f(r)dt^2 - f^{-1}(r)dr^2 - r^2d\theta^2 -r^2\sin^{2}\theta d\phi^2,
\end{equation}
where $f(r) = 1 - 2M/r - \Lambda r^2/3 + (Q_{e}^2 + Q_{m}^2)/r^2 - \alpha r^{-1-3\omega}$, with $M$, $Q_{e}$, and $Q_{m}$ being the mass and the electric and magnetic charges, respectively, $\Lambda$ is the cosmological constant, and $\alpha$ and $-1<\omega<-1/3$ constants associated with the quintessence or dark energy term \cite{kiselev}. Thus, in order to introduce rotational effects into this metric, we employ the Newman--Janis algorithm (NJA) in a manner analogous to that used to obtain rotating black hole solutions \cite{newmanjannis1,newmanjannis2}. Therefore, we first impose that a photon moves radially, such that $d\theta^2 + \sin^{2}\theta d\phi^2 = 0$, and that the trajectory is null, $ds^2 = 0$. We then obtain
\begin{equation}
f(r) d t^2\left[1-f^{-2}(r) \frac{d r^2}{d t^2}\right]=0,
\end{equation}
thus, since $f(r) \neq 0$, we write $dt = f^{-1}(r)\,dr$, and by integrating we obtain
\begin{equation}
t =  \int f^{-1}(r) dr + C,
\end{equation}
where $C$ is a constant, and linearizing we obtain
\begin{equation}\label{tempobarra}
	\bar{t}= t + r - \int f^{-1}(r) dr + C = r + C.
\end{equation}

Thus, differentiating (\ref{tempobarra}), we have $d t = d\bar{t} + \left[f^{-1}(r)-1\right] dr$ then, by squaring both sides, we obtain
\begin{equation}
d t^2 = d \bar{t}^2 + 2 [f^{-1}(r) - 1] d\bar{t}dr + [f^{-1}(r) - 1]^2 d r^2,
\end{equation}
 and substituting $dt^2$ into the metric (\ref{metricageralf}), we obtain
\begin{eqnarray}
		d s^2&=& f(r) d \bar{t}^2+2 [1- f(r)] d \bar{t} d r - \left[2 - f(r)\right] d r^2  -r^2 d \theta^2-r^2 \sin ^2 \theta d \phi^2, 
\end{eqnarray}

Finally, performing the substitution
\begin{equation}
	u=\bar{t}-\int h(r) d r \quad \rightarrow \quad d \bar{t}=d u+h(r) d r,
\end{equation}
we have
\begin{equation}
	d \bar{t}^2 = d u^2+2 h(r) d u d r+h^2(r) d r^2,
\end{equation}
thus, the metric will be
\begin{eqnarray}
		d s^2 &=& f(r) d u^2+ 2 f(r)[h(r)+ f^{-1}(r) - 1] d u d r-f(r)[2f^{-1}(r) - 1 -h^2(r)- \nonumber \\\\  && 2 (f^{-1}(r)-1) h(r)] d r^2  -r^2 d \theta^2-r^2 \sin ^2 \theta d \phi^2.\nonumber
\end{eqnarray}

Continuing with the NJA, we wish to write the metric in Eddington--Finkelstein coordinates, so that the term proportional to $dr^2$ inside the brackets vanishes. Therefore,
\begin{equation}
		h^2(r) + 2 (f^{-1}(r) - 1) h(r)-2f^{-1}(r) +1 = 0,
\end{equation}
so $h(r)=-[(f^{-1}(r) - 1] \pm f^{-1}(r)$. Recall that, as a boundary condition, we want to recover metrics that are already known through this method. Therefore, we choose the positive sign for $h(r)$, so that $h(r)=1$, since in this way we obtain rotating black hole–type metrics (the Kerr metric), for example, when $f(r)= 1-2M/r$. With this, we obtain
\begin{equation}\label{EFmetric}
	ds^2 = f(r)du^2 + 2 dudr - r^2 d\theta^2 - r^2\sin^{2}\theta d\phi^2.
\end{equation}

Thus, having written the metric in (\ref{metricageralf}) in Eddington–Finkelstein coordinates, we can use the NJA to introduce rotation into the metric. To this end, we begin by defining the metric (\ref{EFmetric}) in terms of null vectors, with the metric tensor given by
\begin{equation}\label{tmvetornulo}
	g_{\mu\nu} = l_{\mu}n_{\nu} + l_{\nu}n_{\mu} - m_{\mu}\bar{m}_{\nu} - m_{\nu}\bar{m}_{\mu},
\end{equation}
and
\begin{equation}\label{inversatmvetornulo}
	g^{\mu\nu} = l^{\mu}n^{\nu} + l^{\nu}n^{\mu} - m^{\mu}\bar{m}^{\nu} - m^{\nu}\bar{m}^{\mu},
\end{equation}
which must satisfy the conditions $l_{\mu} l^{\mu}=m_{\mu} m^{\mu}=n_{\mu} n^{\mu}=0$, $l_{\mu} n^{\mu} =-m_{\mu} \bar{m}^{\mu}=1$ and $l_{\mu} m^{\mu} =n_{\mu} m^{\mu}=0$. The metric tensor in (\ref{EFmetric}) and its inverse are given by
\begin{equation}\label{tensormetrico}
	g_{\mu\nu} = \left(\begin{array}{cccc}
		f(r) & 1 & 0 & 0\\
		1 & 0 & 0 & 0 \\
		0 & 0 & -r^2 & 0 \\
		0 & 0 & 0 & -r^2\sin^2\theta
	\end{array}\right), \hspace{1cm} g^{\mu\nu} =  \left(\begin{array}{cccc}
		 0 & 1 & 0 & 0\\
		 1 &-f(r) & 0 & 0 \\
		0 & 0 & \displaystyle -r^{-2} & 0 \\
		0 & 0 & 0 & \displaystyle -r^{-2}\sin^{-2}\theta
	\end{array}\right),
\end{equation}
with $g_{\mu\nu}g^{\mu\lambda} = \delta_{\nu}^{\lambda}$. Thus, considering the inverse of the metric tensor $g^{\mu\nu}$, in order to (\ref{inversatmvetornulo}) to represent (\ref{tensormetrico}), a possible choice for the null tetrad vectors is
\begin{eqnarray}\label{vetornuloctv}
	\left. \begin{array}{l}
		l^{\mu} = \left(0,1,0,0\right) =  \delta_{1}^{\mu}\\\\
		n^{\mu} = \displaystyle \left(1, -\frac{f(r)}{2},0,0\right) = \delta_{0}^{\mu} - \frac{f(r)}{2} \delta_{1}^{\mu}\\\\
		m^{\mu} = \displaystyle \frac{1}{\sqrt{2}r}\left(0,0,1,\frac{i}{\sin\theta}\right) = \frac{1}{\sqrt{2}r}\left(\delta_{2}^{\mu} + \frac{i}{\sin\theta}\delta_{3}^{\mu}\right)\\\\
		\bar{m}^{\mu} = \displaystyle \frac{1}{\sqrt{2}r}\left(0,0,1,-\frac{i}{\sin\theta}\right) = \frac{1}{\sqrt{2}r}\left(\delta_{2}^{\mu} - \frac{i}{\sin\theta}\delta_{3}^{\mu}\right)\\\\
			\end{array}\right.,
\end{eqnarray}
where $l_{\mu} = g_{\mu\nu}l^{\nu}$, $n_{\mu} = g_{\mu\nu}n^{\nu}$ and $m_{\mu} = g_{\mu\nu}m^{\nu}$. Proceeding with the NJA, we now complexify the coordinate $r$ we obtain
\begin{eqnarray}\label{vetornuloctvcompl}
	\left. \begin{array}{l}
		l^{\mu} = \delta_{1}^{\mu},\hspace{4.0cm}
		n^{\mu} = \displaystyle   \delta_{0}^{\mu} - \frac{1}{2} f(r)\delta_{1}^{\mu},\\\\
		m^{\mu} = \displaystyle  \frac{1}{\sqrt{2}\bar{r}}\left(\delta_{2}^{\mu} + \frac{i}{\sin\theta}\delta_{3}^{\mu}\right),\hspace{2.0cm}
		\bar{m}^{\mu} = \displaystyle \frac{1}{\sqrt{2}r}\left(\delta_{2}^{\mu} - \frac{i}{\sin\theta}\delta_{3}^{\mu}\right)
	\end{array}
	\right..
\end{eqnarray}

 Now, performing the coordinate transformation we write $x^{\mu} = (u,r,\theta,\phi) = (u'+ia\cos\theta',r'-ia\cos\theta',\theta',\phi')$, where $u'$, $r'$ and $a = J/M$ are real, with $J$ being the angular momentum of black hole, and using the transformation rules $z'^{\mu} = (\partial x'^{\mu}/\partial x^{\nu}) z^{\nu}$ and $z'_{\mu} = g'_{\mu\nu} z'^{\nu}$, with $z^{\mu} = (l^{\mu},n^{\mu},m^{\mu},\bar{m}^{\mu})$ and $z_{\mu} = (l_{\mu},n_{\mu},m_{\mu},\bar{m}_{\mu})$, we obtain
\begin{eqnarray}\label{vetornuloctvsemifinal}
	\left. \begin{array}{l}
		l'^{\mu} =  \delta_{1}^{\mu},\hspace{2cm}
		n'^{\mu} = \displaystyle \delta_{0}^{\mu} - \frac{1}{2} f(r',\theta')\delta_{1}^{\mu},\\\\
		m'^{\mu} = \displaystyle  \frac{1}{\sqrt{2}(r' + ia\cos\theta')}\left[ ia\sin\theta'(\delta_{0}^{\mu} - \delta_{1}^{\mu})+ \delta_{2}^{\mu} + \frac{i}{\sin\theta'}\delta_{3}^{\mu}\right],\\\\
		\bar{m'}^{\mu} = \displaystyle  \frac{1}{\sqrt{2}(r' - ia\cos\theta')}\left[ -ia\sin\theta'(\delta_{0}^{\mu} - \delta_{1}^{\mu})+ \delta_{2}^{\mu} - \frac{i}{\sin\theta'}\delta_{3}^{\mu}\right]
	\end{array} .
	\right.
\end{eqnarray}
 
To calculate the new form of the function $f(r',\theta')$, we must employ the general complexification rule given by
 \begin{equation}\label{complegeral}
 r^{\beta} \rightarrow \frac{{\rm Re}(r)^{1+\beta/2}}{\bar{r}} \frac{{\rm Re}(r)^{1+\beta/2}}{r} = \frac{r'^{2+\beta}}{\Sigma},
 \end{equation} 
where $\Sigma = r'^{2}+a^2\cos^{2}\theta'$, thus we obtain
 \begin{eqnarray}\label{rcomplexo}
 	\left.\begin{array}{l}
 		\displaystyle	r^{-1-3\omega} \rightarrow  \frac{r'^{1-3\omega}}{\Sigma}\hspace{2.0cm}
 		\displaystyle\frac{1}{r} \rightarrow  \frac{r'}{\Sigma} \hspace{2.0cm}
 		\frac{1}{r^2} \rightarrow  \frac{1}{\Sigma}\hspace{2.0cm}
	r^2 \rightarrow \frac{r'^{4}}{\Sigma}
 	\end{array}\right.,
 \end{eqnarray}
so we have $f(r',\theta') = 1 - 2M r'/\Sigma + (Q_{e}^2 + Q_{m}^2)/\Sigma - \alpha r'^{1-3\omega}/\Sigma - (\Lambda/3)r'^4/\Sigma $. Therefore, we obtain
\begin{equation}\label{tmEF}
	g'^{\mu\nu} = \frac{1}{\Sigma}\left(\begin{array}{cccc}
		\displaystyle -a^2\sin^{2}\theta' & \displaystyle r'^2 + a^2 & 0 & \displaystyle -a\\
		\displaystyle r'^2 + a^2 & \displaystyle -f(r',\theta')\Sigma - a^2\sin^{2}\theta' & 0 & \displaystyle a \\
		0 & 0 & \displaystyle -1 & 0 \\
		\displaystyle -a & \displaystyle a & 0 & \displaystyle -\frac{1}{\sin^{2}\theta'}
	\end{array}\right),
\end{equation}
with the metric tensor
\begin{equation}\label{tm}
	g'_{\mu\nu} = \left(\begin{array}{cccc}
		f(r',\theta') & 1 & 0 & a [1-f(r',\theta')] \sin^{2}\theta'\\
		1 & 0 & 0 & -a\sin^{2}\theta' \\
		0 & 0 & -\Sigma & 0 \\
		a[1-f(r',\theta')] \sin^{2}\theta' & -a\sin^{2}\theta' & 0 & -\sin^2\theta'[\Sigma + a^2 [2-f(r',\theta')]\sin^{2}\theta']
	\end{array}\right),
\end{equation}
and the line element written in generalized Eddington-Finkelstein coordinates will be
\begin{eqnarray}\label{elelinhaef}
	ds^2 &=& f(r',\theta')du'^2 + 2du'dr' + 2a[1-f(r',\theta')] \sin^{2}\theta' du' d\phi' - 2a \sin^{2}\theta' dr' d\phi' \nonumber \\\\ && - \Sigma d\theta'^2 - \sin^{2}\theta'(\Sigma + a^2 [2-f(r',\theta')]\sin^{2}\theta')d\phi'^2.\nonumber
\end{eqnarray}

An important point is that metrics describing, for instance, rotating black holes are commonly written in Boyer–Lindquist coordinates \cite{gravitation,boyer} --- $(t,r,\theta,\phi)$. Therefore, in order to express the line element in these coordinates, in a way analogous to what was done for the Kerr metric in the previous section, we perform the following substitutions
 \begin{equation}\label{mudcoorbl}
 		du' = dt + s(r',\theta')dr', \hspace{1cm}
 		d\phi' = d\phi + g(r',\theta')dr'
 \end{equation} 
where $s(r',\theta')$ and $g(r',\theta')$ are real functions and generalize the result obtained for the Kerr metric. Thus, by substituting (\ref{mudcoorbl}) into (\ref{elelinhaef}), we obtain
 \begin{equation}\label{funcoesfg}
	\left. \begin{array}{l}
		\displaystyle g(r',\theta') = g(r') =  -\frac{a}{\Delta_{r'}} , \hspace{2.0cm}
		\displaystyle s(r',\theta') = s(r') = -\frac{r'^2 + a^2}{\Delta_{r'}}
	\end{array}
	\right.,
\end{equation} 
and the line element in the Boyer–Lindquist coordinates can be written as,
\begin{eqnarray}\label{elelinhablfinal}
	ds^2 &=& \frac{\Delta_{r} - a^2\sin^{2}\theta}{\Sigma} dt^2 - \frac{\Sigma}{\Delta_{r}}dr^2 - \Sigma d\theta^2 + 2a\left(\frac{r^2 + a^2 - \Delta_{r}}{\Sigma}\right)\sin^{2}\theta dt d\phi \nonumber \\\\ && - \frac{\sin^{2}\theta}{\Sigma}\left[(r^2 + a^2)^2  -a^2 \Delta_{r} \sin^{2}\theta\right]d\phi^2,\nonumber
\end{eqnarray}
where $\Delta_{r} = r^{2} + a^2 - 2M r + Q_{e}^{2} + Q_{m}^{2} - \alpha r^{1-3\omega} - \Lambda r^{4}/3$ and for the sake of notational simplicity, from now on we will consider $(r',\theta') \equiv (r,\theta)$.  The metric tensor and its inverse associated with the line element (\ref{elelinhablfinal}) are given by
\begin{equation}\label{tmfinalbl}
	g_{\mu\nu} = \left(\begin{array}{cccc}
		\displaystyle \frac{\Delta_{r} - a^2\sin^{2}\theta}{\Sigma} & 0 & 0 & \displaystyle a\left(\frac{r^2 + a^2 - \Delta_{r}}{\Sigma}\right)\sin^{2}\theta\\
		 0 & \displaystyle -\frac{\Sigma}{\Delta_{r}} & 0 & 0 \\
		0 & 0 & -\Sigma & 0 \\
		\displaystyle a\left(\frac{r^2 + a^2 - \Delta_{r}}{\Sigma}\right)\sin^{2}\theta & 0 & 0 & \displaystyle - \frac{\sin^{2}\theta}{\Sigma}\left[(r^2 + a^2)^2  -a^2 \Delta_{r} \sin^{2}\theta\right]
	\end{array}\right),
\end{equation}
and
\begin{equation}\label{tminversofinalbl}
	g^{\mu\nu} = \left(\begin{array}{cccc}
		\displaystyle \frac{(r^2 + a^2)^2  -a^2 \Delta_{r} \sin^{2}\theta}{\Delta_{r} \Sigma} & 0 & 0 & \displaystyle\frac{a (r^2 + a^2 - \Delta_{r})}{\Delta_{r} \Sigma}\\
		0 & \displaystyle -\frac{\Delta_{r}}{\Sigma} & 0 & 0 \\
		0 & 0 & \displaystyle -\frac{1}{\Sigma} & 0 \\
 \displaystyle\frac{a (r^2 + a^2 - \Delta_{r})}{\Delta_{r} \Sigma}	 & 0 & 0 & \displaystyle -\frac{\Delta_{r} - a^2\sin^{2}\theta}{\Delta_{r} \Sigma \sin^{2}\theta}
	\end{array}\right).
\end{equation}
 
It is interesting to note that the result obtained in (\ref{elelinhablfinal}) possesses some physical and geometrical characteristics distinct from the cases previously investigated for the Kerr--Newman--AdS black hole surrounded by quintessence \cite{xu,xu1,toledo}, as well as from the vacuum case without quintessence \cite{carter}. In those cases, an AdS factorization is introduced into the Kerr--Newman--AdS black hole in order to guarantee a vacuum solution of Einstein's equations, which leads to coupling terms between the cosmological constant $\Lambda$ and the rotation parameter $a$. This differs from the case analyzed here, since we are considering the presence of matter from the very beginning, and in the next section we will calculate the Einstein and stress--energy tensors for this situation. 

Another interesting point is that, starting from the more general result in (\ref{elelinhablfinal}), we are able to recover several other particular black hole solutions. Thus, in the limit $a = 0$, we obtain the result for the charged cosmological black hole surrounded by quintessence with vanishing angular momentum, whose line element is given by
\begin{eqnarray}\label{elelinhablazero}
	ds^2 &=& \frac{\Delta_{r}}{r^2} dt^2 - \frac{r^2}{\Delta_{r}}dr^2 - r^2 d\theta^2 - r^2\sin^{2}\theta   d\phi^2,\nonumber
\end{eqnarray}
where $\Delta_{r} = r^2 - 2Mr + Q_{e}^2 + Q_{m}^{2} - \alpha r^{1-3\omega} - \Lambda r^4/3$.  For other black hole cases with $a\neq 0$, the line element remains the same as that given in (\ref{elelinhablfinal}), with only the function $\Delta_{r}$ being modified. As examples, we have that when $\alpha = 0$, we obtain the rotating charged cosmological black hole with $\Delta_{r} = r^2 - 2Mr + Q_{e}^2 + Q_{m}^{2} - \Lambda r^4/3$; when $\Lambda = 0$, we obtain the rotating charged black hole surrounded by quintessence with $\Delta_{r} = r^2 - 2Mr + Q_{e}^2 + Q_{m}^{2} - \alpha r^{1-3\omega}$ and for $\Lambda = \alpha = 0$, we recover the Kasuya-Kerr-Newman black hole \cite{kasuya}. Finally, when $Q_{e} = Q_{m} = 0$, we obtain the rotating cosmological black hole surrounded by quintessence with $\Delta_{r} = r^2 - 2Mr - \Lambda r^4/3 - \alpha r^{1-3\omega}$.

Finally, using (\ref{tmvetornulo}) and (\ref{inversatmvetornulo}), we can write down the null tetrads associated with the metric tensor and its inverse as follows
\begin{eqnarray}\label{vetornulocovfinal}
	\left. \begin{array}{l}
		\displaystyle l_{\mu} = \left(1,-\frac{\Sigma}{\Delta_{r}},0,-a\sin^{2}\theta\right), \hspace{2.0cm}
		n_{\mu} = \displaystyle \left(\frac{\Delta_{r}}{2\Sigma}, -\frac{1}{2},0,-\frac{a\Delta_{r} \sin^{2}\theta}{2\Sigma}\right), \\\\
		m_{\mu} = \displaystyle \frac{1}{\sqrt{2}}\left(\frac{ia\sin\theta}{r+ia\cos\theta},0,-(r-ia\cos\theta),-\frac{i(r^2 + a^2)\sin\theta}{r+ia\cos\theta}\right),\\\\
		\bar{m}_{\mu} = \displaystyle \frac{1}{\sqrt{2}}\left(-\frac{ia\sin\theta}{r-ia\cos\theta},0,-(r+ia\cos\theta),\frac{i(r^2 + a^2)\sin\theta}{r-ia\cos\theta}\right)
	\end{array}\right.,
\end{eqnarray}
and
\begin{eqnarray}\label{vetornuloctvfinal}
	\left. \begin{array}{l}
		l^{\mu} = \displaystyle \left(\frac{r^2 + a^2}{\Delta_{r}},1,0,\frac{a^2 \Sigma \Delta_{r}^{-1}}{ \Delta_{r} - a^2\sin^{2}\theta}\right) ,\hspace{0.8cm}
		n^{\mu} = \displaystyle \left(\frac{r^2 + a^2}{2\Sigma}, -\frac{\Delta_{r} (\Delta_{r}- a^2\sin^{2}\theta)}{2\Sigma^2},0,\frac{a}{2\Sigma}\right), \\\\
		m^{\mu} = \displaystyle \frac{1}{\sqrt{2}(r+ia\cos\theta)}\left(ia\sin\theta,0,1,\frac{i}{\sin\theta}\right), \\\\
		\bar{m}^{\mu} =\displaystyle \frac{1}{\sqrt{2}(r-ia\cos\theta)}\left(-ia\sin\theta,0,1,-\frac{i}{\sin\theta}\right).
	\end{array}\right. 
\end{eqnarray}

 We concluded the calculation of a rotating charged cosmological black hole surrounded by quintessence. Now, we need to determine how the electromagnetic tensor $F_{\mu\nu}$ is modified due to the curved spacetime with rotational effect. In the next section, we will calculate the electromagnetic and stress-energy tensors.

\subsection{Event horizon}

Returning to the line element (\ref{elelinhablfinal}), we note that the function $\Delta_{r} = \Delta_r(r)$ depends only on the radial coordinate $r$. Thus, by applying the event horizon condition for this black hole, we obtain
\begin{eqnarray}\label{deltafunction}
	\Delta_{r} &=& r^2 -2Mr - \Lambda r^{4}/3 - \alpha r^{1-3\omega} + Q_{e}^{2} + Q_{m}^{2} + a^2  \nonumber\\ &=& (r-r_{-})(r-r_{+})(r-r_{c})(r-r_{q}) = 0,
\end{eqnarray}
which implies that the spherical symmetry is preserved and that the event horizon region will be influenced by all the energy parameters of the system. A forma de $\Delta_r$ definida na segunda linha em (\ref{deltafunction}) é valida somente para valores de $\omega$ que fornecem valores inteiros e positivos para $1-3\omega$.  Furthermore, $r_{-}$ and $r_{+}$ correspond to the inner and Schwarzschild horizons, respectively, since we recover $r_{-} = 0$ and $r_{+} = 2M$ when $\Lambda = \alpha = Q_{e}= Q_{m} = a = 0$. Finally, $r_{c}$ and $r_{q}$ are the cosmological and quintessence horizons, respectively. We also observe that the values of the event horizons are modified according to the value of $\omega$, that is, according to how the quintessence dark energy manifests itself. To better visualize the values of the event horizons and their dependence on different values of the cosmological constant and quintessence parameters, we plot the graphs of $\Delta_{r}$ for different values of $\Lambda$, $\alpha$, and $\omega$. For simplicity, we set $Q_e = Q_m = Q$ in all figures.

\graphicspath{{figuras/}}

\begin{figure}[!htb]
	\centering
	\includegraphics[scale={0.7}]{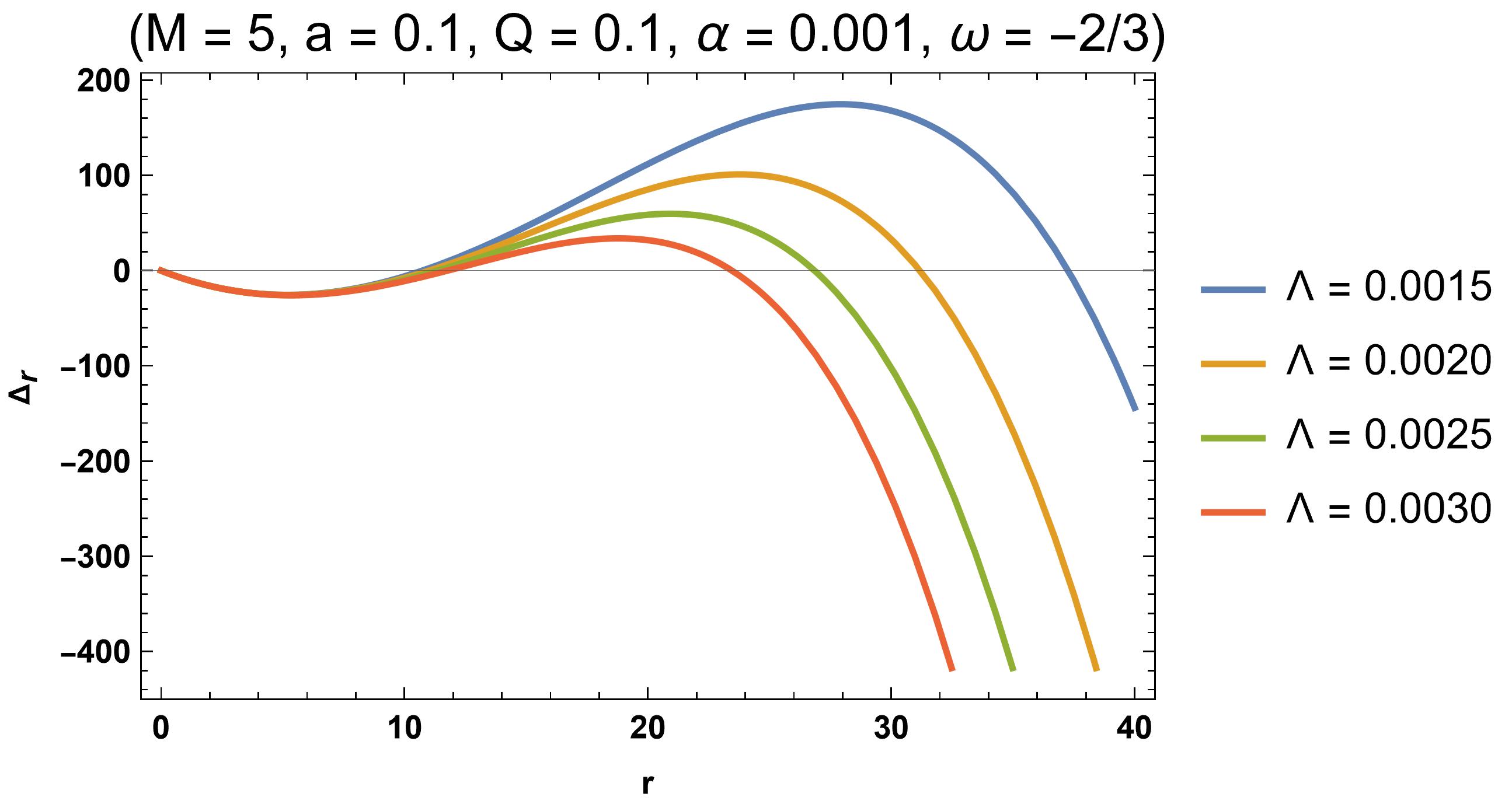}
	\caption{Representation of the behavior of the function $\Delta_{r}$ as a function of $r$ for different values of $\Lambda$.}
	\label{figurardeltar1}
\end{figure}

\graphicspath{{figuras/}}

\begin{figure}[!htb]
	\centering
	\includegraphics[scale={0.7}]{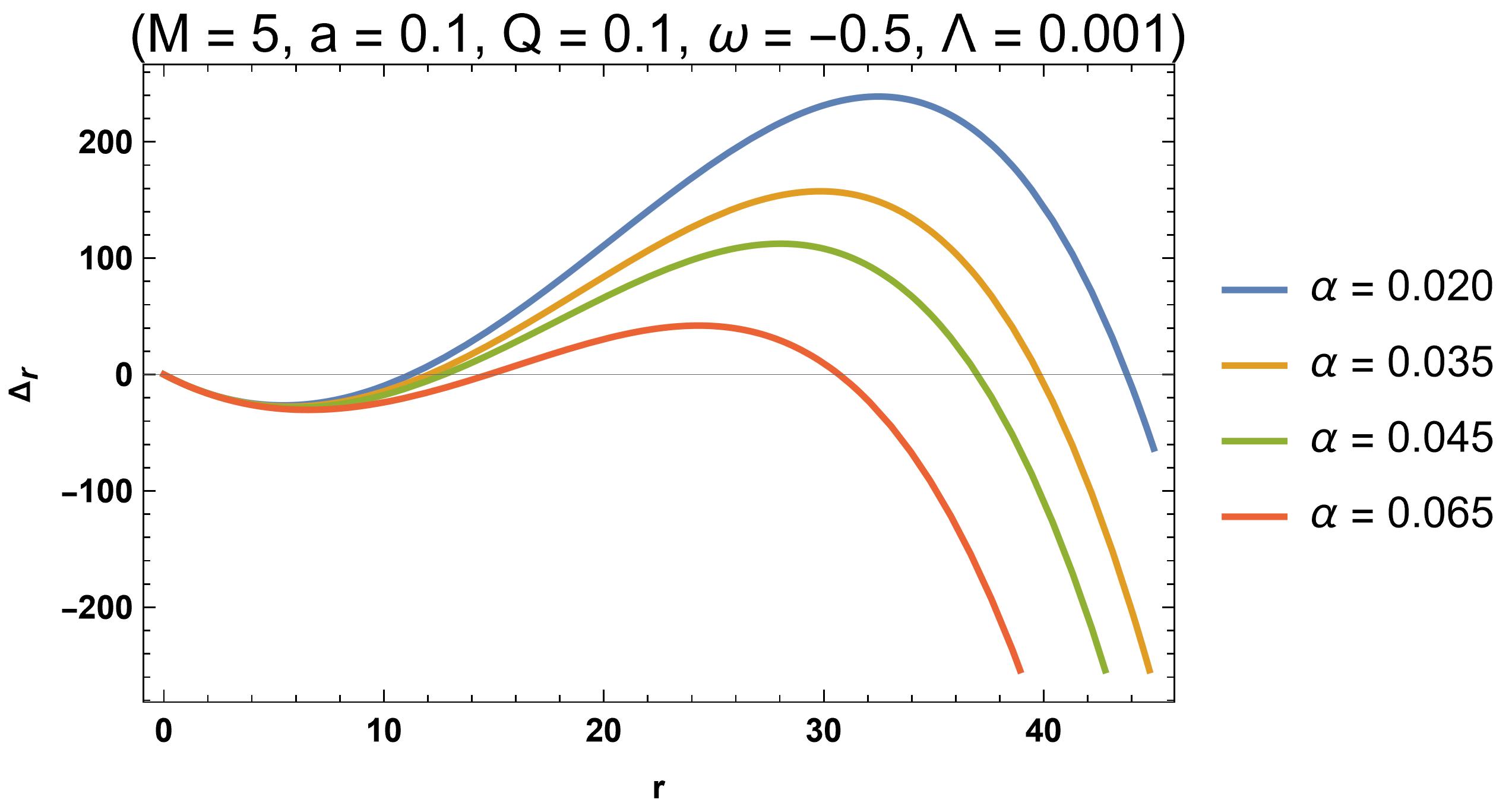}
	\caption{Representation of the behavior of the function $\Delta_{r}$ as a function of $r$ with $\omega = -1/2$ for different values of $\alpha$.}
	\label{figuradeltar2}
\end{figure}

\graphicspath{{figuras/}}

\begin{figure}[!htb]
	\centering
	\includegraphics[scale={0.7}]{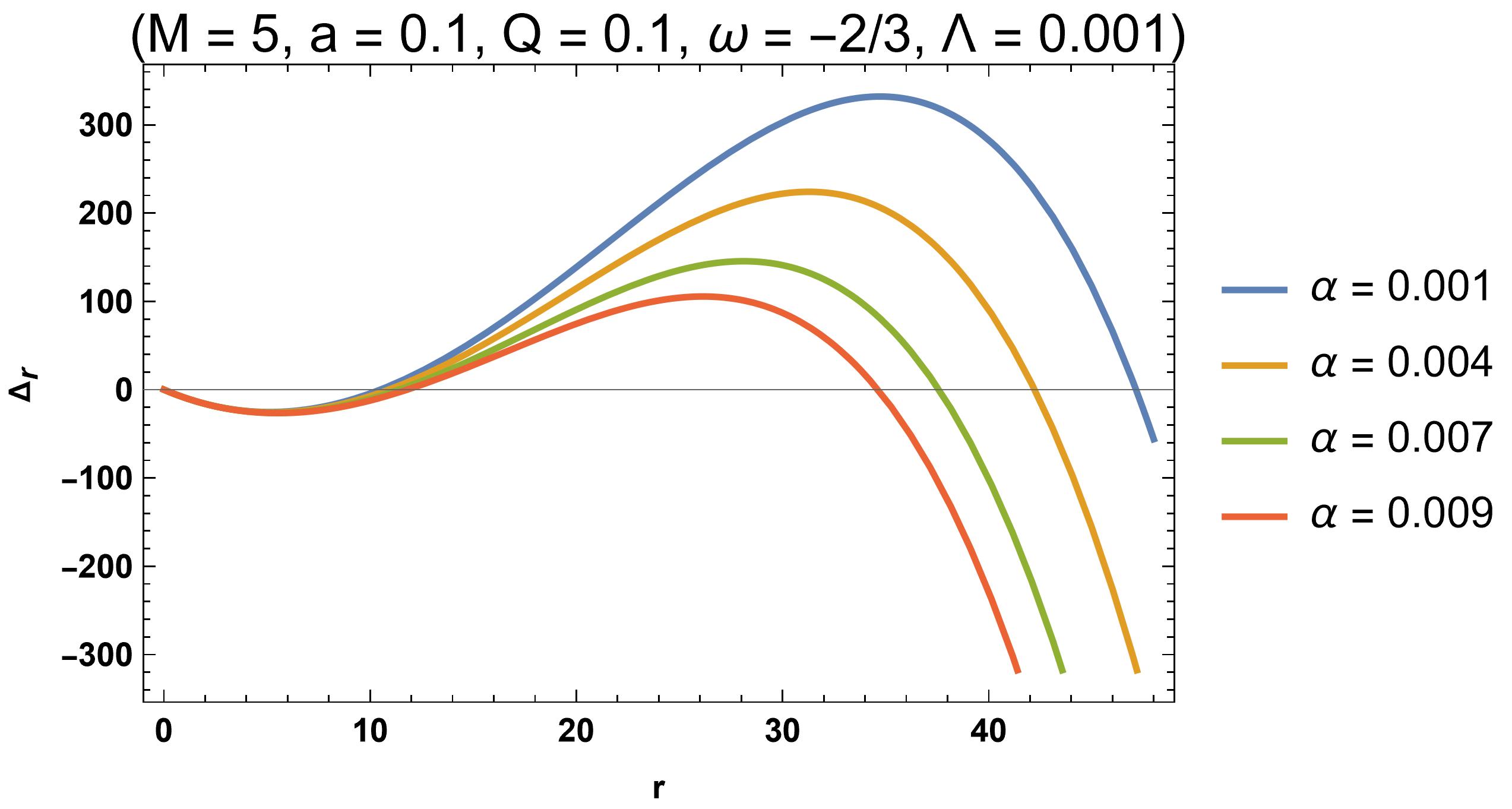}
	\caption{Representation of the behavior of the function $\Delta_{r}$ as a function of $r$ with $\omega = -2/3$ for different values of $\alpha$.}
	\label{figuradeltar3}
\end{figure}

We observe in Figs. (\ref{figurardeltar1}--\ref{figuradeltar3}) that, in all cases, there are only three real values for the event horizons, since the fourth one assumes a complex value. Furthermore, we note that the influence of the cosmological constant $\Lambda$ and quintessence through $\alpha$ and $\omega$ on the values of the event horizons is similar. Thus, the larger the values of $\Lambda$ and $\alpha$, the smaller the values of the horizons.

\subsection{Ergosphere}

\begin{figure}[!htb]
	\centering
	\includegraphics[scale={0.7}]{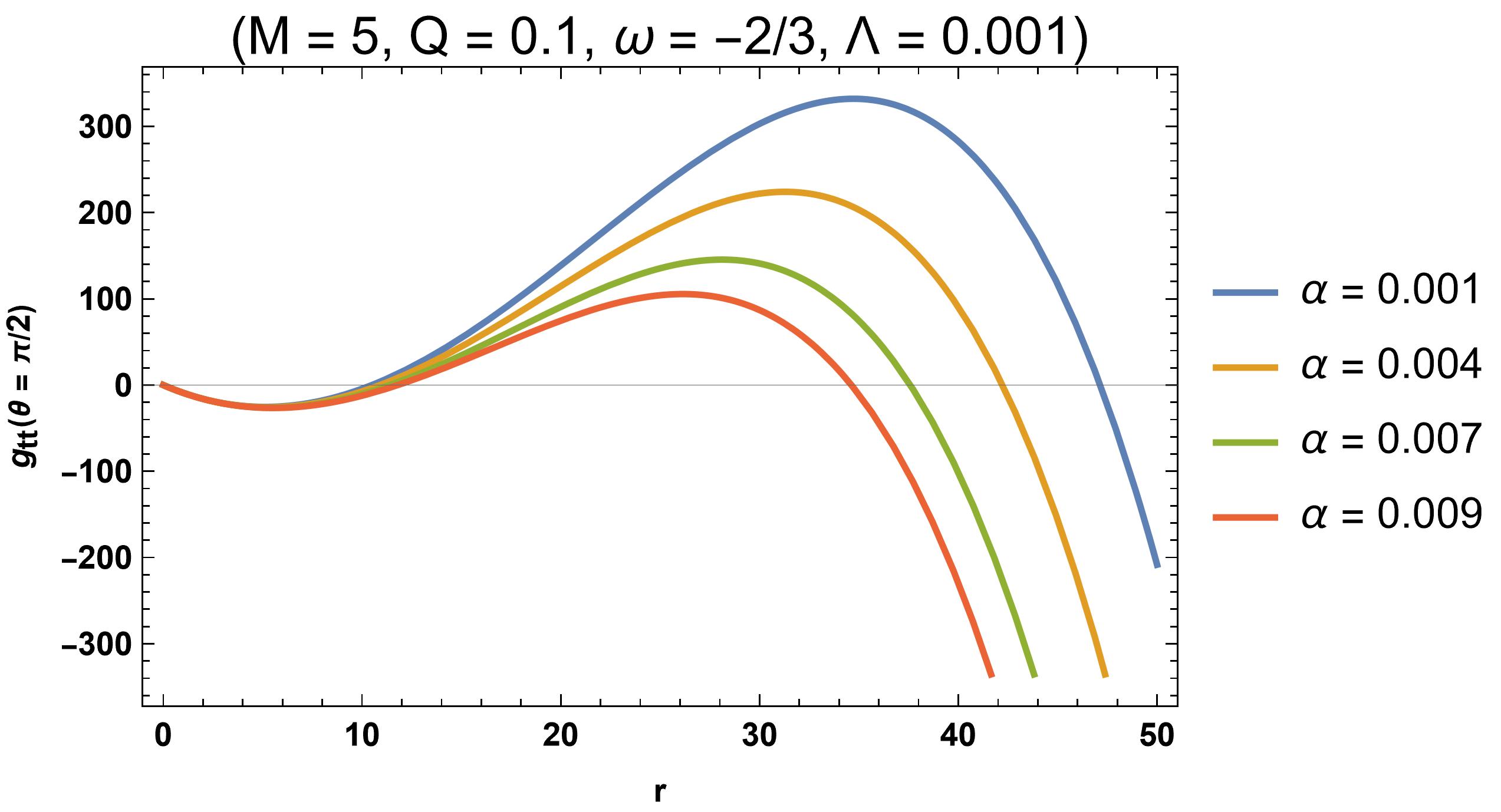}
	\caption{Representation of the behavior of the function $g_{tt}$ as a function of $r$ with $\theta = \pi/2$ and $\omega = -2/3$ for different values of $\alpha$.}
	\label{figuraergo}
\end{figure}

Another region of interest in rotating black holes is the one where a test particle cannot remain free from the dragging effects caused by rotation. This region is called the ergosphere and lies between the event horizon and the region where the particle is no longer influenced by the black hole. To obtain this static surface, or ergosphere, we impose $g_{tt} = 0$, thus 
\begin{equation}
	\Delta_{r} - a^2\sin^{2}\theta = 0 \hspace{1cm} \rightarrow \hspace{1cm} \frac{\Lambda}{3} r^4 - r^2 + 2Mr +\alpha r^{1-3\omega} -Q_{e}^2 - Q_{m}^2 - a^2\cos^{2}\theta = 0,
\end{equation} 
we notice that the static surfaces depend on the parameters $\Lambda$, $M$, $\alpha$, $\omega$, and $a$, and this equation can be solved numerically. A particular case occurs in the equatorial plane, when $\theta = \pi/2$, where we now obtain static circumferences given by
\begin{equation}
	\frac{\Lambda}{3} r^4 - r^2 + 2Mr +\alpha r^{1-3\omega} -Q_{e}^2-Q_{m}^2  = (r-r_1)(r-r_2)(r-r_3)(r-r_4) = 0,
\end{equation}  
therefore one obtain four values of $r$ for the static circumferences, and these values are independent of $a$. In Fig. (\ref{figuraergo}), we plot the function $g_{tt}$ for $\theta = \pi/2$ in order to better visualize the values of $r$ corresponding to the static circumferences for different values of $\alpha$, for example with $\omega = -2/3$. We again consider $Q_{e} = Q_{m} = Q$.

\graphicspath{{figuras/}}

We note that Fig. (\ref{figuraergo}) exhibits behavior similar to that observed in the figures from the previous section, which was already expected since $g_{tt}(\theta = \pi/2) = \Delta_{r} - a^2$. Thus, the larger the value of $\alpha$, the smaller the values of the static circumference. In particular, comparing with Fig. (\ref{figuradeltar3}), since we used the same parameter values in both figures, for the event horizon value $r_{h}$ obtained from $\Delta_{r=r_{h}} = 0$, we impose that the static circumference satisfies $r_{sc} > r_{h}$, where $g_{tt}(\theta = \pi/2) = 0$ at $r_{sc}$.

\section{Electromagnetic and Stress-Energy Tensors }

In this section, we first calculate the electromagnetic tensor $F^{0}_{\mu\nu}$ for the static black hole case whose metric is given in (\ref{metricageralf}), and then include the effects of angular momentum using the null vectors obtained in the previous section. Here, the superscript “0” denotes the case without rotational effect. Therefore, let the Maxwell equations be given by
\begin{eqnarray}\label{maxwelleq}
	\nabla_{\beta}(F^{\alpha\beta}) = 0, \\
	\nabla_{\gamma}(F_{\alpha\beta}) + \nabla_{\alpha}(F_{\beta\gamma}) + \nabla_{\beta}(F_{\alpha\gamma}) = 0,
\end{eqnarray} 
with $\nabla_{\beta}(F^{\alpha\gamma}) = \partial_{\beta}F^{\alpha\gamma} + \Gamma_{\mu\beta}^{\alpha}F^{\mu\gamma} + \Gamma_{\mu\beta}^{\gamma}F^{\alpha\mu}$ where $\Gamma_{\gamma\mu}^{\nu}$ is the Christoffel symbol,
\begin{equation}
	\Gamma_{\gamma\mu}^{\nu} = \frac{1}{2}g^{\nu\lambda}(\partial_{\mu}g_{\gamma\lambda} + \partial_{\gamma}g_{\lambda\mu} - \partial_{\lambda}g_{\mu\gamma}),
\end{equation}
being symmetric in the lower indices $\Gamma^{\sigma}_{\mu\nu} = \Gamma^{\sigma}_{\nu\mu}$ and the tensor $F^{0}_{\mu\nu}$ will be given by
\begin{equation}
	F^{0}_{\mu\nu} = \left(\begin{array}{cccc}
		0 & -F^{0}_{01} & 0 & 0 \\
		F^{0}_{10} & 0 & 0 & 0 \\
		0 & 0 & 0 & -F^{0}_{32}\\
		0 & 0 & F^{0}_{23} & 0 
	\end{array}\right),
\end{equation}
with $F^{0\mu\nu} = g^{\mu\alpha}g^{\nu\beta}F^{0}_{\alpha\beta}$. As we want to compute the tensor $F^{0}_{\mu\nu}$ in the metric (\ref{metricageralf}), the metric tensor $g_{\mu\nu}$ and its inverse $g^{\mu\nu}$ are given by
\begin{eqnarray}
	g_{\mu\nu} = {\rm diag}\left[f(r),-f^{-1}(r),-r^2,-r^2\sin^{2}\theta\right], \\\nonumber\\
	g^{\mu\nu} = {\rm diag}\left[f^{-1}(r),-f(r),-\frac{1}{r^2},-\frac{1}{r^2\sin^{2}\theta}\right],
\end{eqnarray} 
and the nonvanishing Christoffel symbols are
\begin{eqnarray}
	\left. \begin{array}{r}
	\displaystyle \Gamma^{1}_{00} = \frac{f(r)}{2}\frac{df(r)}{dr}, \hspace{1.2cm} \Gamma^{1}_{11} = - \frac{1}{2f(r)}\frac{df(r)}{dr},\hspace{1.2cm} \Gamma^{1}_{22} = -r f(r), \\\\
	\displaystyle  \hspace{1.2cm} \Gamma^{1}_{33} = -r f(r)\sin^{2}\theta, \hspace{1.0cm} \Gamma^{2}_{12} = \Gamma^{2}_{21} = \frac{1}{r},\hspace{1.2cm} \Gamma^{2}_{33} = -\sin\theta\cos\theta \\\\
	\displaystyle  \Gamma^{3}_{13} = \Gamma^{3}_{31} = \frac{1}{r}, \hspace{1.2cm} \Gamma^{3}_{23} = \Gamma^{3}_{32} = \cot\theta, \hspace{1.2cm} \Gamma^{0}_{10} = \Gamma^{0}_{01} = \frac{1}{2f(r)}\frac{df(r)}{dr}.
		\end{array}
	\right.,
\end{eqnarray}
where $f(r) = 1 - 2M/r - \Lambda r^2/3 + (Q_{e}^2 + Q_{m}^2)/r^2 - \alpha r^{-1-3\omega}$.

Therefore, due to spherical symmetry we have $F_{01} = -F_{10} =  E_{r}(r)$ and $F_{23} = -F_{32} = B_{r}(r)r^2\sin\theta$, since $B_{\alpha} = g_{\alpha\beta}/\sqrt{|g|},\epsilon^{0\beta\mu\nu} F_{\mu\nu}$, where $E_{r}$ and $B_{r}$ are the radial components of the electric and magnetic fields, respectively, $\epsilon^{0\beta\mu\nu}$ is the Levi-Civita symbol, and $g = {\rm det}(g_{\mu\nu})$. Thus, using (\ref{maxwelleq}), we obtain
\begin{eqnarray}
		\displaystyle E_{r}(r) = E(r) = \frac{Q_{e}}{r^2}\hspace{1.5cm} {\rm and} \hspace{1.5cm}	B_{r}(r) = B(r) = \frac{Q_{m}}{r^2}
\end{eqnarray}
with that, $F_{23} = -F_{32} = Q_{m}\sin\theta$. After obtaining the electromagnetic tensor in the static metric (\ref{metricageralf}), we can now include the effect of rotation in this tensor. To this end, we will use the null tetrads obtained in (\ref{vetornulocovfinal}) and (\ref{vetornuloctvfinal}), the values of $F_{01}$, $F_{10}$, $F_{23}$ and $F_{32}$ obtained above, and the Maxwell scalars \cite{chandra} defined by
\begin{eqnarray}
		\phi_{0} = F_{\mu\nu}l^{\mu}m^{\nu},\hspace{1.5cm}
		\phi_{1} = \frac{1}{2}F_{\mu\nu}(l^{\mu}n^{\nu} + \bar{m}^{\mu}m^{\nu}),\hspace{1.5cm}
		\phi_{2} = F_{\mu\nu}\bar{m}^{\mu}n^{\nu},
\end{eqnarray}
such that we obtain $\phi_{0} = \phi_{2} = 0$ and $\phi_{1} = (Q_{e} + iQ_{m})/(2r^2)$. Thus, the new electromagnetic tensor in the black hole with angular momentum is defined by $F_{\mu\nu} = 2\phi_{1}(n_{\mu}l_{\nu} + m_{\mu}\bar{m}_{\nu}) + c.c. $, where the second term is the complex conjugate of the first one. Using the complexification rules in $\phi_{1}$ with $r \rightarrow r-ia\cos\theta$ we obtain $\phi_{1} = (Q_e + iQ_m)/2\Sigma$, and carrying out the calculations, we find that the nonvanishing terms of the new electromagnetic tensor, written in the $2$-form, are given by
\begin{equation}
	{\bf F} = F_{10}\; {\bf d}r\wedge dt + F_{13}\; {\bf d}r\wedge {\bf d}\phi + F_{20}\; {\bf d}\theta \wedge {\bf d}t + F_{23}\; {\bf d}\theta \wedge {\bf d}\phi ,
\end{equation}
or written more explicitly
\begin{eqnarray}\label{tensorem}
	{\bf F} &=&  -\frac{\sin\theta}{\Sigma^2}[2Q_{e}ar\cos\theta + Q_{m}(r^2-a^2\cos^{2}\theta)]{\bf d}\theta \wedge [a{\bf d}t - (r^2 + a^2){\bf d}\phi] \nonumber\\\\ && + \frac{1}{\Sigma^2}[Q_{e}(r^2-a^2\cos^{2}\theta)-2Q_{m}ar\cos\theta]{\bf d}r\wedge ({\bf d}t - a\sin^{2}\theta {\bf d}\phi),\nonumber
\end{eqnarray}
the electromagnetic field above continues to satisfy the Maxwell equations in (\ref{maxwelleq}), as expected. The electromagnetic stress--energy tensor in the signature convention adopted here is defined by \cite{gravitation}
\begin{equation}\label{SEeletroma}
	T_{\mu\nu}^{\rm EM} =  \frac{1}{4\pi}\left(\frac{1}{4}g_{\mu\nu}F^{\alpha\beta}F_{\alpha\beta}  -  g^{\alpha\beta}F_{\mu\alpha}F_{\nu\beta}  \right).
\end{equation}

An important point is that, in the case where the gravitational object contains additional energy components that deform spacetime beyond the electric charge, the total stress–energy tensor will include extra contributions in addition to the electromagnetic one. Finally, we now need to determine the form of the Ricci tensor $R_{\mu\nu}$, which is given by\cite{gravitation}
\begin{equation}
	R_{\mu\nu} = \partial_{\sigma}\Gamma^{\sigma}_{\mu\nu} - \partial_{\nu}\Gamma^{^\sigma}_{\mu\sigma} + \Gamma^{\sigma}_{\sigma \lambda}\Gamma^{\lambda}_{\mu\nu} - \Gamma^{\sigma}_{\mu\lambda}\Gamma^{\lambda}_{\sigma\nu},
\end{equation}
so that the metric tensor in (\ref{tmfinalbl}) satisfies Einstein’s equation. Thus, the total stress-energy tensor is given by \cite{gravitation}
\begin{equation}\label{stressenergia}
	T_{\mu\nu} = \frac{1}{k}\left[R_{\mu\nu} - \frac{R}{2}g_{\mu\nu}\right] = \frac{1}{k} G_{\mu\nu} ,
\end{equation}
where $R = g^{\mu\nu}R_{\mu\nu}$ is the scalar curvature, $G_{\mu\nu}$ is the Einstein tensor, $k$ an constant given by $k = 8\pi G/c^4$. 

Therefore, in order to determine the components of $T_{\mu\nu}$ for the metric (\ref{tmfinalbl}), we must first compute the Ricci tensor and then the Einstein tensor. Performing these calculations, we obtain that the nonvanishing components of the tensor $T_{\mu\nu} = (1/k)G_{\mu\nu}$ are given by
\begin{eqnarray}\label{einsteintensor}
	\displaystyle T_{tt} &=& -\frac{3\alpha \omega r^{-1-3\omega}}{k\Sigma^3}\left[r^2\Delta_r - a^4\sin^2\theta\cos^2\theta + \frac{a^2}{2}(3\omega+1)\Sigma\sin^2\theta\right] +\frac{(Q_e^{2}+Q_m^{2})}{k\Sigma^3}\times\nonumber \\\\&& (\Delta_r + a^2\sin^2\theta) + \frac{\Lambda r^2}{k\Sigma^3}[r^2\Delta_r -a^2\sin^2\theta (r^2+2a^2\cos^2\theta)] \nonumber,
\end{eqnarray}
\begin{eqnarray}
	 \displaystyle T_{rr} = -\frac{3\alpha \omega}{k\Sigma \Delta_r} r^{1-3\omega} - \frac{Q_e^{2} + Q_m^{2}}{k\Sigma \Delta_r} - \frac{\Lambda r^4}{k\Sigma \Delta_r} ,
\end{eqnarray}	 
\begin{eqnarray}
	\displaystyle T_{\theta\theta} &=& \frac{3\alpha \omega r^{-1-3\omega}}{k\Sigma}\left(a^2\cos^2\theta - \frac{(3\omega+1)}{2}\Sigma\right) - \frac{(Q_e^{2} + Q_e^{2})}{k\Sigma} - \frac{\Lambda r^2}{k\Sigma}(r^2 + 2a^2\cos^2\theta),
\end{eqnarray}
\begin{eqnarray}	
	T_{t\phi} &=& \frac{3a\alpha\omega r^{-1-3\omega} \sin^{2}\theta }{k\Sigma^3}[2(r^2(\Delta_r - a^4\cos^2\theta) + (1+3\omega)(r^2+a^2)\Sigma ] + \frac{a(Q_e^{2} + Q_m^{2})}{k\Sigma^3}\sin^{2}\theta \times \nonumber\\\\ &&(r^2+a^2 + \Delta_r) 
	+ \frac{\Lambda a\sin^{2}\theta r^2}{k\Sigma^3}[-2(r^2(\Delta_r - a^4\cos^2\theta) + 2(r^2+a^2)\Sigma] \nonumber,
\end{eqnarray}
\begin{eqnarray}	
	T_{\phi\phi} = \frac{3\alpha\omega r^{-1-3\omega} \sin^{2}\theta}{k\Sigma^3} \left\{a^2[(r^2+a^2)^2\cos^2\theta-r^2\Delta_r \sin^2\theta] - \frac{(1+3\omega)}{2}(r^2+a^2)^2\Sigma \right\} \nonumber \\\\ - \frac{(Q_e^{2}+ Q_m^{2})}{k\Sigma^3}[(r^2+a^2)^2 + a^2\Delta_r \sin^2\theta]\sin^{2}\theta - \frac{\Lambda r^2 \sin^{2}\theta}{k\Sigma^3}[a^2 + (r^2+a^2)^2\Sigma]\nonumber
\end{eqnarray}

An interesting feature is that the components of the total stress--energy tensor $T_{\mu\nu}$, computed above, contain the energy contributions associated with quintessence, the electric and magnetic charges, and the cosmological constant, which are proportional to the parameters $\alpha$, $Q_e^{2} + Q_m^{2}$, and $\Lambda$, respectively. Furthermore, the electromagnetic stress--energy tensor contributions are identical to those given in Eq.~(\ref{SEeletroma}). Thus, we conclude the calculation of the electromagnetic tensor and the Einstein and total stress-energy tensors. As expected, the obtained values of $G_{\mu\nu}$ and $T_{\mu\nu}$ differ from those found for the vacuum Kerr--Newman--AdS black hole case \cite{xu} in both situations, namely when $\alpha = 0$ and $\alpha \neq 0$, since here we consider the spacetime in the presence of matter.

\subsection{Singularities}

Another important characteristic of black holes is the possible presence of singularities. To examine this, we use the Kretschmann scalar, which for rotating Schwarzschild-type metrics is typically given by
\begin{equation}\label{kretschmannscalar}
	K = R_{\mu\nu\sigma\rho}R^{\mu\nu\sigma\rho} = \frac{H(a,\Lambda,\omega,Q_e,Q_m,r,\theta)}{\Sigma^{12}},
\end{equation}
where $R_{\mu\nu\sigma\rho}$ is the Riemann tensor, and $H(a,\Lambda,\omega,Q_e,Q_m,r,\theta)$ is a function that depends on the black hole parameters and the coordinates $r$ and $\theta$.
The singularity exists when $K \rightarrow \infty$, that is, when $\Sigma = 0$. Thus, we conclude that the singularity is located at $r = 0$ and $\theta = \pi/2$, analogously to the vacuum case \cite{xu}. 

Moreover, considering the Ricci scalar given by
\begin{equation}\label{ricciscalar}
	R = g^{\mu\nu}R_{\mu\nu} = -\frac{r^2[4\Lambda - 3\alpha \omega (1-3\omega)r^{-3-3\omega}]}{\Sigma},
\end{equation}
we observe that both rotation and dark energy directly influence the energy associated with the cosmological constant, since by setting $a = \alpha = 0$, we recover the value $R = -4\Lambda$, corresponding to the presence of a cosmological constant in vacuum.

\subsection{Rotational and angular velocities}

Now, we calculate the rotational and angular velocities of a unit-mass test particle. For this purpose, we again restrict the motion to the equatorial plane with $\theta = \pi/2$ and $d\theta/dt = 0$. Another important point is that we investigate the particle motion in a locally non-rotating frame (LNRF), that is, the angular momentum vanishes in this reference frame. Thus, the Lagrangian of the system is given by \cite{oteev}
\begin{eqnarray}
	\mathcal{L} &=& \frac{1}{2} g_{\alpha\beta}\frac{\partial x^{\alpha}}{\partial \tau}\frac{\partial x^{\beta}}{\partial \tau}\nonumber\\\\
	&=& \frac{1}{2}\left[g_{tt} \left(\frac{\partial t}{\partial \tau}\right)^2 + g_{rr} \left(\frac{\partial r}{\partial \tau}\right)^2 + 2g_{\mu\nu} \left(\frac{\partial t}{\partial \tau}\right) \left(\frac{\partial \phi}{\partial \tau}\right) + g_{\phi\phi} \left(\frac{\partial \phi}{\partial \tau}\right)^2\right],\nonumber
\end{eqnarray}
where $\tau$ is the proper time in the static reference frame and $u^{\mu} = \partial x^{\mu}/\partial \tau = \dot{x}^{\mu}$ are the four-velocities. So, the associated momenta are given by

\begin{eqnarray}\label{momentosassociados}
	\left\{\begin{array}{l}
		p_{t} = g_{tt} \dot{t} + g_{t\phi} \dot{\phi} = -E \\
		p_{r} = g_{rr}\dot{r}\\
		p_{\phi} = g_{t\phi} \dot{t} + g_{\phi\phi}\dot{\phi} = L
	\end{array}\right.,
\end{eqnarray}
where $E$ and $L$ are the mechanical energy and angular momentum of the test particle, respectively. The Hamiltonian is given by $\mathcal{H} = \frac{1}{2}(p_{t}\dot{t} + p_{r}\dot{r} + p_{\phi}\dot{\phi}) - \mathcal{L}$, and imposing $2\mathcal{H} = -1$, since we analyze the particle motion along a timelike geodesic \cite{oteev}, we obtain
\begin{eqnarray}
	\dot{r}^2 = -\frac{1}{g_{rr}} + \frac{g_{\phi\phi}E^2 + 2g_{t\phi}EL + g_{tt}L^2}{g_{rr}(g_{tt}g_{\phi\phi} - g_{t\phi}^2)} =  (E-V_{eff})(E+V_{eff}),
\end{eqnarray}
where $V_{eff}$ is an effective potential and we use $\dot{t} = -(g_{\phi\phi}E + g_{t\phi}L)/(g_{tt}g_{\phi\phi} - g_{t\phi}^2)$ and $\dot{\phi} = (g_{tt}L + g_{t\phi}E)/(g_{tt}g_{\phi\phi} - g_{t\phi}^2)$ obtained through (\ref{momentosassociados}).  Applying the restriction of spherical symmetry with circular orbits, we have $\dot{r} = 0$, consequently $V_{eff} = E$ and $dV_{eff}/dr = 0$. Thus, by applying these two conditions, we obtain, as in \cite{oteev}
\begin{eqnarray}
	L &=& \pm \frac{g_{t\phi} + g_{\phi\phi}\Omega_{\phi}}{\sqrt{-g_{tt}-g_{t\phi}\Omega_{\phi}-g_{\phi\phi}\Omega_{\phi}^2}} \nonumber\\\\ &=&
	\pm \frac{a(r^2+a^2 - \Delta_r) + [a^2\Delta_r - (r^2+a^2)^2]\Omega_{\phi}}{r\sqrt{a^2-\Delta_r + 2(\Delta_r - r^2-a^2)\Omega_{\phi} - (a^2\Delta_r - (r^2+a^2)^2)\Omega_{\phi}^{2}}}\nonumber
	\end{eqnarray}
	\begin{eqnarray} E &=& \pm \frac{g_{tt} + g_{t\phi}\Omega_{\phi}}{\sqrt{-g_{tt}-g_{t\phi}\Omega_{\phi}-g_{\phi\phi}\Omega_{\phi}^2}} \nonumber \\\\ &=& 
		\pm \frac{\Delta_r - a^2 + a^2(r^2+a^2 - \Delta_r)\Omega_{\phi}}{r\sqrt{a^2-\Delta_r + 2(\Delta_r - r^2-a^2)\Omega_{\phi} - (a^2\Delta_r - (r^2+a^2)^2)\Omega_{\phi}^{2}}},\nonumber
\end{eqnarray}
where the rotational velocity is given by
\begin{eqnarray}
	v &=& \frac{L}{\sqrt{g_{\phi\phi}}} = \pm \frac{1}{\sqrt{g_{\phi\phi}}}\frac{g_{t\phi} + g_{\phi\phi}\Omega_{\phi}}{\sqrt{-g_{tt}-g_{t\phi}\Omega_{\phi}-g_{\phi\phi}\Omega_{\phi}^2}}\nonumber\\\\ &=& \pm
	\frac{a(a^2 + r^2-\Delta_r) + [a^2\Delta_r - (r^2+a^2)^2]\Omega_{\phi}}{\sqrt{[a^2\Delta_r - (r^2+a^2)^2][a^2-\Delta_r + 2(\Delta_r - r^2-a^2)\Omega_{\phi} - (a^2\Delta_r - (r^2+a^2)^2)\Omega_{\phi}^{2}]}} \nonumber .
\end{eqnarray}
and the angular velocity $\Omega_{\phi}$ is
\begin{eqnarray}
	\Omega_{\phi} &=& \frac{-g_{t\phi,r} \pm \sqrt{(g_{t\phi,r})^2 + g_{tt,r}g_{\phi\phi,r}}}{g_{\phi\phi,r}} \nonumber \\\\
	&=& \frac{a(2a^2 -2\Delta_r + r\Delta'_{r}) \pm \sqrt{(2\Delta_r - r\Delta'_r - 2a^2) [2a^2(2\Delta_r - r\Delta'_{r}-2a^2) + 2r^4]}}{a^2(2a^2 - 2\Delta_r - r\Delta'_{r}) - 2r^4} .\nonumber
\end{eqnarray}

It is interesting to note that even in the absence of angular momentum, with $a = 0$, the values of the rotational and angular velocities remain nonzero due to the locally non-rotating reference frame considered in our calculations. Furthermore, we observe that all energy terms influence the physical quantities calculated above, and when $\alpha = \Lambda = 0$, we recover the corresponding values for the Kerr--Newman black hole \cite{oteev}.

\subsection{Required matter and energy conditions}

According to Einstein's theory of General Relativity, the matter distribution associated with a spacetime geometry must satisfy a set of physical requirements known as the energy conditions, namely the null, weak, strong, and dominant conditions \cite{hawking,gravitation}. In order to analyze whether these requirements are fulfilled in the present rotating black hole spacetime, we investigate the corresponding stress--energy tensor components. However, because the metric contains non-diagonal terms related to rotation, the usual expressions for the energy conditions in diagonal geometries cannot be directly employed. To properly perform this analysis, we consider the observer in a locally nonrotating frame (LNRF) \cite{bardeen,bambi}, where the line element assumes the form
\begin{equation}
	ds^2 = g_{\mu\nu}dx^{\mu}dx^{\nu} = e^{a}e^{b}\eta_{ab},
\end{equation}
with the basis vectors given by $e^{a} = e^{a}_{\mu}dx^{\mu}$ where $e^{a}_{\mu}$ are the tetrads which satisfy the ortogonality condition $e^{a}_{\mu}e^{\mu}_{b} = \delta^{a}_{b}$ and $\eta_{ab} = {\rm diag}(1,1,1,1)$, with $(\mu,\nu) = (t,r,\theta,\phi)$ and $(a,b) = (0,1,2,3)$. Thus, for the line element given in (\ref{elelinhablfinal}), the basis vectors are given by
\begin{eqnarray}
	\left.\begin{array}{l}
	\displaystyle	{\bf e}^{0} = e^{0}_{t}{\bf dt} = \left|g_{tt} - \frac{g_{t\phi}^2}{g_{\phi\phi}}\right|^{1/2}{\bf dt},\hspace{2cm}
		{\bf e}^{1} = e^{1}_{r}{\bf dr} =  |g_{rr}|^{1/2}{\bf dr},\\\\
		{\bf e}^{2} =e^{2}_{\theta}{\bf d\theta} =  |g_{\theta\theta}|^{1/2}{\bf d\theta}, \hspace{2cm}
	\displaystyle	{\bf e}^{3} = e^{3}_{t}{\bf dt} + e^{3}_{\phi}{\bf d\phi} = \frac{g_{t\phi}}{|g_{\phi\phi}|^{1/2}}{\bf dt} + |g_{\phi\phi}|^{1/2}{\bf d\phi}
	\end{array}\right.,
\end{eqnarray}
the absolute values used above are introduced solely to avoid complex values arising from the square roots. When these terms are squared, we obtain $(e^{i}_{\mu})^2 = g_{\mu\mu}$. In this way, the stress--energy tensor $T^{\mu\nu}$ can be rewritten in the orthonormal basis \cite{poisson}
\begin{equation}
	T^{\mu\nu} = \epsilon e^{\mu}_{0}e^{\nu}_{0} + p_{1}e^{\mu}_{1}e^{\nu}_{1} + p_{2}e^{\mu}_{2}e^{\nu}_{2} + p_{3}e^{\mu}_{3}e^{\nu}_{3},
\end{equation}
where $\epsilon = T^{\mu\nu}e^{0}_{\mu}e^{0}_{\nu}$ is the energy density, and $p_{i} = T^{\mu\nu}e^{i}_{\mu}e^{i}_{\nu}$ represent the principal pressures of the fluid. Using (\ref{stressenergia}) and $T^{\mu\nu} = g^{\alpha\mu}g^{\beta \nu} T_{\alpha\beta}$, we obtain
\begin{eqnarray}
\epsilon = \frac{8a^2\sin^2\theta \Delta_r^{2} - 8(r^2+a^2)^2(a^2-r^2+ r\Delta'_r) + \beta \Delta_r}{8k \Sigma^2\xi},
\end{eqnarray}
\begin{equation}
	p_{1} = \frac{2r^2}{k\Sigma \Delta_{r}}\rho'  \hspace{3cm}
	p_{2} = \frac{2a^2\cos^{2}\theta}{k\Sigma}\rho' + \frac{r}{k}\rho'',
\end{equation}
\begin{eqnarray}
	p_{3} = \left(\frac{1}{8k\Delta_r\Sigma^4}\right)\left[2a^2\sin^2\theta(r^2+a^2-\Delta_r)\lambda + 2\xi\chi + \frac{a^2\sin^2\theta(r^2+a^2-\Delta_r)^2}{\xi}\times\right.\nonumber \\\\ \left.  \left(-8a^2\sin^2\theta\Delta_r^{2} + 8(r^2+a^2)^2(a^2-r^2+r\Delta_r) - \xi \Delta_r\right) \right],\nonumber
\end{eqnarray}
where 
\begin{eqnarray}
	\left. \begin{array}{l}
		\xi = (r^2+a^2)^2 - a^2\Delta_r \sin^2\theta\\\\
		\beta = 3a^4 + 16a^2r^2 + 8r^4 + a^4(4\cos2\theta + \cos4\theta) + 2a^2\sin^2\theta(2\Sigma \Delta''_r - 4r\Delta'_r)\\\\
		\lambda = -4a^2 + 4r^2+4(r^2 - a^2\cos^2\theta +\Delta_r)\Delta_r - 4r(r^2+a^2+\Delta_r)\Delta'_r + 2\Sigma \Delta_r \Delta''_r\\\\
		\chi = -4\Delta_r^{2} + 4a^2\sin^2\theta(a^2 - r^2 + r\Delta'_{r}) + \Delta_r (8a^2\cos^2\theta + 4r\Delta'_{r} - 2\Sigma \Delta''_r)
	\end{array}\right.,
\end{eqnarray}
with $\rho = (\alpha r^{-3\omega} + 2M - Q^2/r + \Lambda r^3/3)/2$.

Therefore, within the LNRF frame, we consider, for instance, that the conditions corresponding to the lowest energy cases are given by 
\begin{eqnarray}
	\epsilon + p_{i}\geq 0, \hspace{1cm} i=(1,2,3), \hspace{1cm}\text{ for null energy condition}\\
	\epsilon\geq 0 \hspace{0.5cm}{\rm and}\hspace{0.5cm}\epsilon + p_{i}\geq 0, \hspace{1cm} i=(1,2,3), \hspace{1cm}\text{ for weak energy condition}.
\end{eqnarray}

Restricting once again the analysis to the equatorial plane, with $\theta = \pi/2$, we observe that all parameters of the system directly affect the energy density $\epsilon$, the pressures $p_i$, and consequently the energy conditions. We find that whenever the null energy condition is violated, all remaining energy conditions are violated as well. Furthermore, there are two situations in which the energy density and pressures diverge for specific values of $r$: when $\Delta_r = 0$, corresponding to the event horizon region $r_h$, and when $\xi = 0$, which defines a region distinct from the event horizon. However, the values satisfying $\xi = 0$ are always located in the region $r < r_h$, where the spacetime may present between one and four real values for $r_h$. Therefore, the region defined by $\xi = 0$ is not physically relevant for our analysis, since we are interested only in the spacetime outside the event horizon.

As an example of a black hole configuration satisfying the weak energy conditions above, we consider the case $\omega = -2/3$, $Q = 0.1$, $M = 1$, $\alpha = 0.1$, $\Lambda = 0.001$, and $a = 1$, for $r \in (2.58,7.06)$, where $r_h \approx 1.70$. Another configuration that satisfies the weak energy conditions is obtained when $\omega = -1/2$, $Q = 0.1$, $M = 1$, $\alpha = 0.1$, $\Lambda = 0.001$, and $a = 1$, with $r \geq 2.32$, where $r_h = 1.530$.

Therefore, when the energy conditions are satisfied, the spacetime can be supported by non-exotic matter, for which every observer measures a positive energy density. This corresponds to the expected behavior of classical matter. On the other hand, when one or more energy conditions are violated, the geometry requires the presence of exotic matter or fields, such as the quintessence considered in the present model, or even quantum effects \cite{wald,kontou}.

\section{Conclusion}

In this work, we constructed the metric of a rotating cosmological black hole with electric and magnetic charges surrounded by quintessence, considering the Einstein field equations in the presence of external matter. Starting from the line element (\ref{metricageralf}) and the function
$f(r) = 1 - 2M/r - \Lambda r^2/3 + (Q_{e}^2 + Q_{m}^2)/r^2 - \alpha r^{-1-3\omega}$, we applied the Newman--Janis algorithm (NJA) to introduce rotation. We also established the general complexification rule given in (\ref{complegeral}), removing ambiguities in the procedure described in (\ref{rcomplexo}).

Due to the symmetry of the system, the electromagnetic tensor obtained in (\ref{tensorem}) is independent of $\Lambda$ and $\alpha$. Using the resulting metric, we calculated the stress--energy tensor $T_{\mu\nu} = G_{\mu\nu}/k$, and the scalar curvature (\ref{ricciscalar}), investigating the effects of external matter on the spacetime geometry. Unlike the result obtained in \cite{xu}, where the Kerr--Newman--AdS black hole was considered in vacuum, our solution explicitly includes matter associated with the cosmological constant. In the limit $\Lambda \rightarrow 0$, we recover the dyonic Kerr--Newman black hole surrounded by quintessence.

We found that the function $\Delta_r$ contains no interaction terms between $\Lambda$ and $a$. The figures (\ref{figurardeltar1}--\ref{figuradeltar3}) show that increasing $\Lambda$ and $\alpha$ reduces the real values of the event horizons. A similar behavior is observed for the ergosphere, since $g_{tt}(\theta = \pi/2) = \Delta_r - a^2$. The physical singularity occurs at $r = 0$ and $\theta = \pi/2$, where $R \rightarrow \infty$ and $K \rightarrow \infty$, in agreement with \cite{xu}.

Using the metric tensor in the LNRF frame, we calculated the angular and rotational velocities, as well as the mechanical energy and angular momentum of a test particle. These quantities are directly influenced by the energy terms present in the model, and the Kerr--Newman results are recovered when $\alpha = \Lambda = 0$. We also analyzed, within this same framework, the energy conditions through the energy density $\epsilon$ and pressures $p_i$. We observe that the violation of the weak energy condition implies the violation of all other energy conditions. Divergences in the energy density and pressures occur at $\Delta_r = 0$, corresponding to the event horizon $r_h$, and at $\xi = 0$; however, the latter always satisfies $r < r_h$, and therefore does not affect the region outside the horizon. Depending on the parameter choice, the energy conditions may be satisfied in some external regions, indicating non-exotic matter and positive energy density, while in other cases they are violated throughout the exterior region, suggesting the presence of exotic matter or fields, such as quintessence, as well as possible quantum effects \cite{wald,kontou}.

Finally, in the limit $a = 0$, we recover the electrically and magnetically charged cosmological black hole surrounded by quintessence \cite{kiselev}. For cases with $a \neq 0$, the line element remains unchanged, with only the function $\Delta_r$ being modified. In particular, when $\alpha = 0$, we obtain a rotating electrically and magnetically charged cosmological black hole; when $\Lambda = 0$, we recover the dyonic Kerr--Newman black hole surrounded by quintessence; and for $Q_{e} = Q_{m} = 0$, we obtain the rotating cosmological black hole surrounded by quintessence.

\section*{Acknowledgements}
AGMS and MDO gratefully acknowledges CNPq (grant number 309052/2023-8) and FAPERJ (grant number 200.247/2026) for partial financial support. This study was funded by FAPERJ --- Fundação Carlos Chagas Filho de Amparo à Pesquisa do Estado do Rio de Janeiro, Process SEI 26/200.337/2024.

\end{document}